\documentclass[preprint,aps]{revtex4}
\usepackage{graphicx}
\begin{document}
\title{Solar arcades as possible minimum dissipative relaxed
states}
\author{R. Bhattacharya$^1$, M. S. Janaki$^1$, B. Dasgupta$^2$ and G.
P. Zank$^2$}

\affiliation{$^1$ Saha Institute of Nuclear Physics, I/AF
Bidhannagar, Calcutta
700 064, India\\
$^2$ Institute of Geophysics and Planetary Physics, University of
California at Riverside, Riverside, CA 92521, USA}

\email{dasgupta@ucr.edu}

\begin{abstract}
Arcade-type magnetic field structures originating from the
photosphere are relevant to the understanding of different types
of solar prominences, coronal loops and coronal heating. In most
of the existing literature, these loop-like magnetic structures
are modeled as force-free fields (FFF) without any plasma flow.
The system is assumed to be isolated and non-dissipative. In
reality the photospheric plasma hardly qualifies to be isolated in
nature and the existence of an external drive in the form of a
dynamo field is always possible. Solar prominences are not ideal
either since dissipative effects are believed to play a major role
in coronal heating. The above observations indicate that a
force-free model based on a non-dissipative plasma may not be a
suitable candidate to replicate the arcade structures and further
investigations are required. In our work, we have proposed the
arcade structures as minimum dissipative relaxed states (including
both the viscous and resistive channels) pertaining to a two-fluid
description of the plasma. The obtained relaxed state is non
force-free in nature and appropriate to an open system with
external drives. The Euler-Lagrange equations are solved in
Cartesian coordinates subject to the existing photospheric
boundary conditions. The solutions are seen to support
flow-containing arcade like magnetic field configurations with
inherent dissipative properties that may play an important role in
the coronal heating. An interesting feature observed is the
generation of different types of arcades with the variation of a
single parameter characterizing the relaxed state.  Also, recent
observations with the LASCO coronagraph on board the SOHO
spacecraft suggest that the helmet streamers originating from the
sun may have an internal triple-arcade structure. The two-fluid
relaxed state obtained here is also seen to support such
structures.
\end{abstract}

\maketitle

\section{Introduction}

Arcade type magnetic structures play an essential role in solar
physics, particularly regarding the coronal loops, coronal heating
and coronal mass ejection. A typical coronal loop consists of a
bunch or a single arch-like magnetic field line(s) joining two
magnetically active regions of opposite polarity on the
photosphere \cite{kn:krishan1, kn:priest1}. The active regions
themselves are separated by a neutral line, where the normal
component of the magnetic field is zero. In general the coronal
loops are observed to be \emph{long-lived}. By \emph{long-lived}
we mean the life-time of the structure involved is more than that
expected from a conventional magnetohydrodynamic (MHD) analysis.
The coronal loops are seen to have life-times varying from an hour
to more than one day \cite{kn:foukal1} whereas the MHD time scale
for a typical loop parameter is of the order of seconds
\cite{kn:chiuderi} only.

The global structure of the magnetic arcades are modelled either
by assuming a potential (current-free) field \cite{kn:poletto} or
by considering the plasma to be force-free
 \cite{kn:priest2, kn:browning,
kn:aly1, kn:bellanpaper} and the magnetic field satisfying the
equation

\begin{equation}
\nabla\times{\bf{B}}=\lambda{\bf{B}}
\end{equation}

\noindent with $\lambda$ as a spatial function or a constant. The
fff model is  justified by an an order of magnitude analysis that
shows the structures to be of low $\beta$ \cite{kn:nakagawa} or on
arguments similar to that of Taylor in the context of laboratory
plasmas \cite{kn:taylor1, kn:taylor2}, where it is envisioned as a
relaxed state characterized by the magnetic energy minimum along
with an invariant magnetic helicity \cite{kn:priest2,
kn:browning}. The second scheme appears to be more lucrative,
since by definition it makes the arcade structures long-lived and
at the same time imparts a sense of stability to them. This stems
from the fact that the relaxed state being a spontaneously
achieved state, any deviation from it is expected to generate the
same mechanisms that were responsible to attain the relaxed state
at the very first place. As a result, the system will once again
try to converge to its earlier state and the perturbations will
die out. In other words, there is always an anticipation that the
relaxed states are also inherently stable by nature. But this is
only an expectation and a detail analysis \cite{kn:linden1,
kn:linden2} is always required to study the relevant stability
properties.

Although the force-free description is successful in explaining
the gross topological features of the magnetic arcades but at the
same time also suffers from major drawbacks. As observed through
EUV spectrum \cite{kn:foukal1} the plasma temperature has a
monotonically increasing profile inside the loop being minimum at
the axis and maximum at the surface. The density shows no
significant variation transverse to the loop axis
\cite{kn:chiuderi} and even can be considered to be uniform
\cite{kn:foukal2}. These observations imply that the pressure is
not an uniform parameter inside the loop and a force-free
description may not be an appropriate candidate to model the
arcade. In a recent work by G. A. Gary \cite{kn:allen1} it has
been recognized that the solar corona itself has regions of
different plasma $\beta$. It is only the mid-corona sandwiched
between the photosphere at the base of the coronal loop and the
upper-corona where the plasma $\beta<<1$ and the force-free
condition can be assumed to hold good. The other two regions are
essentially characterized by $\beta>1$ and hence support non
force-free fields. Based on the SXT Limb Data \cite{kn:allen2},
Gary has proposed a model that admits a range of $\beta$ values in
the mid corona. Also in an alternative scenario the solar flares
are proposed to be natural consequence of the high $\beta$ plasma
confined in a curved coronal loop \cite{kn:shibasaki}. The
above-mentioned works essentially points that a force-free model
is not adequate in explaining all the features of arcade fields
and a non force-free description is highly desirable.

Another feature missing in the force-free description of the
arcade fields is the plasma flow. In all the force-free models
mentioned above,  the plasma is assumed to be ideal or nearly
ideal. This essentially decouples the magnetic and flow fields,
making them independent of each other. In contrast to this finite
plasma flows have been observed in the coronal loops
\cite{kn:krishan2}, a typical form being the Evershed flow which
is directed out of the sunspots at the photosphere and into them
in the chromosphere \cite{kn:surlanzis}. A finite pressure
difference between the foot points is believed to be responsible
for such flows. Flows are important in context of the solar winds
also where the plasma is seen to flow outwards along the open
field lines. The importance of flow has recently been recognized
by Mahajan \emph{et. al.} \cite{kn:mahajan1} also. According to
them viscous dissipation does play an important role as long as
the flow field is treated as an integral part of the plasma
dynamics and accounts for the coronal heating during the very
formation of the loops.

From the above discussions it emerges that although the force-free
model is suitable in explaining the gross topological features of
the arcade-type magnetic fields, further studies must be made in
order to accommodate features such as non-uniform pressure profile
and finite plasma flow. To include such features, in the following
we propose the arcade fields as driven relaxed states governed by
the principle of \emph{minimum dissipation rate} (MDR). Before
going any further, we want to declare that our goal is rather
modest and is confined only to the feasibility study of the arcade
structures as MDR relaxed state. In this paper we will be more
concerned with the general topological features and leave all the
quantitative analysis; like the heating rate or the magnitude of
the plasma flow for future works. Also for analytical progress,
essential in understanding the natural coupling of the
self-consistent magnetic fields and non-zero plasma flows, we will
assume the density to be constant. The situations where such an
assumption is valid being already been discussed by Mahajan
\emph{et. al} \cite{kn:mahajan1}, no further clarifications are
given here.

\section{Formulation of the problem}
To formulate a variational problem for obtaining the relaxed
state, we need a minimizer (minimization of which yields the
relaxed state) and invariants or constraints that essentially
determines how far the system will be propped away from its
trivial outcome. A crucial point to be noted is, most of the
arcade fields, including the solar corona, are externally driven
magnetoplasma systems. For the solar corona, the external drive is
provided by the convective motions in the bounding photospheric
surface penetrated by the coronal magnetic field
\cite{kn:ortolani}. With dissipative processes playing a key role
\cite{kn:mahajan1}, such systems are not in thermodynamic
equilibrium and we have to look for suitable minimizers and
invariants. Principle of minimum dissipation rate first proposed
by Montgomery and Philips in the context of plasma physics
\cite{kn:montgomery} is one of the suitable candidates to describe
such off-equilibrium magnetoplasma systems related to various
fusion oriented configurations \cite{kn:bhattacharyya1,
kn:farengo, kn:zhang} (and the references therein). We must point
out that in absence of a rigorous mathematical proof, principle of
MDR still enjoys the status of a conjecture- the validity of which
depends on the agreement of predictions based on it with the
experimental observations. The success of MDR in predicting
experimentally realizable relaxed states in laboratory plasmas
makes it a viable candidate for the astrophysical plasmas also.

In the following we describe the plasma by two-fluid equations.
This has two basic advantages over its single-fluid or MHD
counterpart. In two-fluid description, at-least in context of the
relaxation physics, plasma flow gets coupled to the magnetic field
in a natural way. Also, the two-fluid description being more
general than MHD, it is expected to capture certain results that
are otherwise unattainable from a single-fluid description. In
two-fluid formalism the magnetic helicity is replaced by the
generalized helicities defined as

\begin{equation}
K_\alpha=\int_V{\bf{P}}_\alpha\cdot{\bf{\Omega}}_\alpha d\tau
\end{equation}

\noindent where ${\bf{P}}_\alpha$ and ${\bf{\Omega}}_\alpha$ are
the canonical momentum and canonical vorticity for the $\alpha$
species and are given by

\begin{eqnarray}
& &{\bf{P}}_\alpha=m_\alpha{\bf{u}}_\alpha+\frac{q_\alpha}{c}{\bf{A}}\nonumber\\
& &\\
& &{\bf{\Omega}}_\alpha=\nabla\times{\bf{P}}_\alpha \nonumber
\end{eqnarray}

\noindent As can easily be acknowledged, the expression for
$K_\alpha$ in equation (2) is gauge independent only if the
boundary condition ${\bf{\Omega}}_\alpha \cdot\hat{n}=0$ is
satisfied. Such a boundary condition makes the system to be
isolated from the surroundings. To accommodate arcade structures,
which are inherently driven systems \cite{kn:ortolani}, we must
modify the above definition of generalized helicity. An
alternative expression for the generalized helicity can be found
by extending the definition of Jensen and Chu \cite{kn:jensen} to
the two-fluid case. Let us define \cite{kn:bhattacharyya2}

\begin{equation}
K_\alpha=\int_V{\bf{P}}_\alpha\cdot{\bf{\Omega}}_\alpha d\tau-
\int_V{\bf{P}}_\alpha^\prime\cdot{\bf{\Omega}}_\alpha^\prime d\tau
\end{equation}

\noindent where ${\bf{P}}_\alpha$ and ${\bf{P}}_\alpha^\prime$ are
different inside the volume of interest and are same outside. The
gauge invariance can easily be seen by writing
${\bf{P}}_\alpha={\bf{P}}_\alpha+\nabla\psi$, $\psi$ being some
arbitrary scalar continuous at the boundary.

\begin{equation}
K_\alpha^\prime=K_\alpha+\oint \psi({\bf{\Omega}}_\alpha
-{\bf{\Omega}}_\alpha^\prime)\cdot\hat{n}da
\end{equation}

\noindent where the surface integral is to be evaluated over the
bounding surface $S$ of the volume $V$. In writing the above
equation we have used the Gauss' theorem along with the fact that
$\nabla\cdot{\bf{\Omega}}_\alpha( {\bf{\Omega}}_\alpha^\prime)=0$.
So the generalized helicity $K_\alpha$ as defined by equation (4)
is gauge invariant provided the boundary condition

\begin{equation}
({\bf{\Omega}}_\alpha-{\bf{\Omega}}_\alpha^\prime) \cdot\hat{n} =0
\end{equation}

\noindent is obeyed. If we further assume that there is no
flow-field coupling at the surface i.e., the flow and the magnetic
fields are two independent variables and the variation of one does
not affect the other, we have

\begin{eqnarray}
& &({\bf{B}}-{\bf{B}}^\prime)\cdot\hat{n}=0 \nonumber\\
& &\\
& &({\bf{\omega}}_\alpha-{\bf{\omega}}_\alpha^\prime)
\cdot\hat{n}=0\nonumber
\end{eqnarray}

\noindent at the surface, where

\begin{equation}
{\bf{\omega}}_\alpha=\nabla\times{\bf{u}}_\alpha
\end{equation}

\noindent is the vorticity. Equations (7) then represent a type of
natural boundary conditions inherent to the problem. Next, with
generalized helicity given by equation (4), in the following we
derive the helicity balance equation.

For the $\alpha$ component the momentum balance equation is given
by,

\begin{equation}
m_\alpha n_\alpha \frac{d{\bf{u}}_\alpha}{dt} =n_\alpha
q_\alpha\left({\bf{E}}+\frac{1}{c}{\bf{u}}_\alpha\times{\bf{B}}\right)-\nabla
h_\alpha +\mu_\alpha\nabla^2{\bf{u}}_\alpha-n_\alpha q_\alpha
\eta{\bf{J}}
\end{equation}

\noindent where we have assumed the flow to be incompressible and
$h_\alpha=p_\alpha+g$, where $g$ is the gravitational potential
and $p$ is the pressure. Writing
${\bf{E}}=-\nabla\phi-(1/c)(\partial{\bf{A}}/{\partial t})$ and
using the vector identity

\begin{equation}
({\bf{u}}_\alpha\cdot\nabla){\bf{u}}_\alpha=\nabla\frac{u_\alpha^2}{2}
-{\bf{u}}_\alpha\times(\nabla\times{\bf{u}}_\alpha)
\end{equation}

\noindent the momentum balance equation can be cast as,

\begin{eqnarray}
\frac{\partial{\bf{P}}_\alpha}{\partial t}&=&-\nabla\left[q_\alpha
\phi+ \frac{h_\alpha}{n_\alpha}+\frac{m_\alpha
u_\alpha^2}{2}\right]
+{\bf{u}}_\alpha\times{\bf{\Omega}}_\alpha \nonumber\\
& &-\frac{\mu_\alpha}{n_\alpha}(\nabla\times)^2{\bf{u}}_\alpha
-\frac{q_\alpha\eta c}{4\pi}(\nabla\times){\bf{B}}
\end{eqnarray}

\noindent where we have used the notation
$(\nabla\times)^n=\nabla\times\nabla\times...$to n terms.
\noindent Taking curl on both sides, we get the vorticity balance
equation as

\begin{equation}
\frac{\partial{\bf{\Omega}}_\alpha}{\partial t}=
\nabla\times({\bf{u}}_\alpha\times{\bf{\Omega}}_\alpha)
-\frac{\mu_\alpha}{n_\alpha}(\nabla\times)^3{\bf{u}}_\alpha
-\frac{q_\alpha\eta c}{4\pi}(\nabla\times)^2{\bf{B}}
\end{equation}

\noindent Using the above two equations along with the definition
(4) for the generalized helicity,  corresponding balance equation
for the $\alpha$ component is obtained as,

\begin{eqnarray}
\frac{d K_\alpha}{dt}&=&-\oint[G_\alpha-{\bf{P}}_\alpha\cdot{\bf
{u}}_\alpha]{\bf{\Omega}}_\alpha^\prime\cdot\hat{n}da-\oint({\bf{P}}_\alpha
\cdot{\bf{\Omega}}_\alpha){\bf{u}}_\alpha\cdot\hat{n}da \nonumber\\
&-&\oint\left(\frac{\partial{\bf{P}}_\alpha^\prime}{\partial t}
 \times{\bf{P}}_\alpha^\prime\right)\cdot\hat{n}da
 -\eta_\alpha^\prime\oint[\{\nabla\times{\bf{B}}+L_\alpha
 {\bf{\omega}}_\alpha)\}\times{\bf{P}}_\alpha]\cdot\hat{n}da \\
 &-&2\eta_\alpha^\prime\int[\nabla\times({\bf{B}}+L_\alpha
 {\bf{\omega}}_\alpha)]\cdot{\bf{\Omega}}_\alpha d\tau \nonumber
 \end{eqnarray}

 \noindent with

 \begin{eqnarray}
 & &G_\alpha=\left[q_\alpha\phi+\frac{h_\alpha}{n_\alpha}-\frac{m_\alpha
 u_\alpha^2}{2}\right] \nonumber\\
 & &\eta_\alpha^\prime=\frac{q_\alpha\eta c}{4\pi} \nonumber\\
 & & \\
 & & L_\alpha=\frac{4\pi}{q_\alpha c
 n_\alpha}\frac{\mu_\alpha}{\eta}\equiv~~{\rm{Prandtl~~number}}\nonumber
\end{eqnarray}

\noindent While deriving the above equation, we have used the
boundary condition described by equation (6). In the above balance
equation the surface integrals represent the helicity injection
terms while the volume integral represents the dissipation rate
for $K_\alpha$. For an isolated system characterized by
${\bf{\Omega}}_\alpha^\prime\cdot\hat{n}=0$,
${\bf{u}}_\alpha\cdot\hat{n}=0$ and $\hat{n}\times{\bf{P}}=0$ (the
last condition is due to the fact that in isolated systems with
conducting material boundary, $\hat{n}\times{\bf{A}}=0$ and
$\hat{n}\times{\bf{u}}_i=0$ which is the no-slip boundary
condition and is applicable to viscous systems), only the
dissipation term in the balance equation  survives and for a
static plasma characterized by ${\bf{u}}_\alpha=0$ is similar to
the dissipation rate of the magnetic helicity.

To proceed further, in the following we consider a two-component
hydrogen plasma. Neglecting the electron mass over the ion mass
with $q_i=-q_e=q$ and the quasineutrality to hold good, we have

\begin{eqnarray}
& &\eta_i^\prime=\frac{q\eta c}{4\pi} \nonumber\\
& &\eta_e^\prime=-\frac{q\eta c}{4\pi} \nonumber\\
& &\\
& & L_i=\frac{4\pi}{q c
 n}\frac{\mu_i}{\eta} \nonumber\\
 & & L_e=0 \nonumber
 \end{eqnarray}

\noindent In writing the last expression we have neglected the
 electron viscosity  over the ion viscosity due to its smallness
 by the electron to ion mass ratio.

 Following Montgomery \emph{et al} \cite{kn:montgomery}, let us
assume that $K$ is a bounded function. Then the time average of
$K$ over a reasonably long time will make left side of the above
equation a constant. Prescribing the surface terms \cite{kn:bevir}
then it is possible to form a variational problem where the
relaxed state is obtained by minimizing the total dissipation
rates, both ohmic and viscous, with the helicity dissipation rates
as constraints. In other words, in our model we have assumed the
field variables to be fixed at the surface. This assumption is
more of a mathematical requirement to keep the subsequent analysis
simple and at per with the variational principle in classical
mechanics. The minimization integral is then obtained as,

\begin{eqnarray}
 {\cal{I}}&=&\int
 \left[(\nabla\times{\bf{B}})\cdot(\nabla\times{\bf{B}})+\frac{4\pi
 n q L_i}{c}{\bf{\omega}}_i\cdot{\bf{\omega}}_i\right]d\tau \nonumber\\
 &+&\lambda_i\int[\nabla\times({\bf{B}}+L_i{\bf{\omega}}_i)]\cdot{\bf{\Omega_i}}
 d\tau-\lambda_e\int(\nabla\times{\bf{B}})\cdot{\bf{\Omega}}_e
 d\tau
 \end{eqnarray}

\noindent the first two terms in the above integral represents the
ohmic and viscous dissipation rates while the second two integrals
represent the ion and electron generalized helicity dissipation
rates. $\lambda_i$ and $\lambda_2$ are the corresponding Lagrange
undetermined multipliers. The Euler-Lagrange equations are
obtained by equating the first order variation of the above
integral to zero and treating $\delta{\bf{B}}$ and
$\delta{\bf{u}}_i$ as independent variations and are given by,

\begin{eqnarray}
&
&(\nabla\times)^2{\bf{B}}+\frac{(\lambda_i+\lambda_e)q}{c}(\nabla\times){\bf{B}}
+\frac{\lambda_i}{2}\left( \frac{q
L_i}{c}+m_i\right)(\nabla\times){\bf{\omega}}_i=0 \\
&
&(\nabla\times)^2{\bf{\omega}}_i+\frac{2\pi n q}{m_i\lambda_i
c}(\nabla\times){\bf{\omega}}_i+\frac{1}{2 m_i L_i}\left( \frac{q
L_i}{c}+m_i\right)(\nabla\times)^2{\bf{B}}=0
\end{eqnarray}

\noindent Eliminating ${\bf{B}}$ (${\bf{u}}_i$) in favor of
${\bf{u}}_i$ $({\bf{B}})$, the EL equations can be rewritten as

\begin{eqnarray}
&
&(\nabla\times)^2{\bf{B}}+\left[\frac{(\lambda_i+\lambda_e)q}{c}+\frac{2\pi
n q}{m_i\lambda_i c}-\frac{\lambda_i}{4 m_i L_i}\left( \frac{q
L_i}{c}+m_i\right)^2\right](\nabla\times){\bf{B}} \nonumber\\
& &+\frac{2\pi n q^2}{m_i
c^2}\frac{\lambda_i+\lambda_e}{\lambda_i}{\bf{B}}=\nabla\psi
\end{eqnarray}

\begin{eqnarray}
&
&(\nabla\times)^2{\bf{\omega_i}}+\left[\frac{(\lambda_i+\lambda_e)q}{c}+\frac{2\pi
n q}{m_i\lambda_i c}-\frac{\lambda_i}{4 m_i L_i}\left( \frac{q
L_i}{c}+m_i\right)^2\right](\nabla\times){\bf{\omega_i}} \nonumber\\
& &+\frac{2\pi n q^2}{m_i
c^2}\frac{\lambda_i+\lambda_e}{\lambda_i}{\bf{\omega_i}}=\nabla\chi
\end{eqnarray}

\noindent where $\psi$ and $\chi$ satisfy Laplace's equation

\begin{equation}
\nabla^2\psi(\chi)=0
\end{equation}

\noindent A few comments are necessary for the above set of
equations. For a constant gauge, the equations reduce to the
double-curl Beltrami equations obtained by Mahajan and Yoshida
\cite{kn:mahajan2}. Also the equations are in some sense robust to
a variation in the Prandtl number. For both the limits
$L_i\rightarrow 0$ and $L_i\rightarrow \infty$ and for finite
$L_i$ the equations maintain their double-curl nature. As the
finite $L_i$ case also includes the ideal limit ($\eta\rightarrow
0$, $\mu_i\rightarrow 0$), this shows that the flow does play an
important role in securing a relaxed as well as a steady-state in
the ideal limit, as conjectured by Montgomery and Philips
\cite{kn:montgomery}.

\section{Solution of the EL equations: Arcade structures}

To obtain solutions of the EL equations pertaining to arcade
structures let us employ the cartesian coordinates. We assume that
the $x-y$ plane represents the photospheric surface and arcade
structures are extended to the positive half $z-$ plane. So our
volume of interest is the region characterized by the positive
values of $z$ coordinate.

To solve the EL equations, let us define

\begin{eqnarray}
& &{\bf{B}}^\dagger={\bf{B}}-\frac{2\pi n q^2}{m_i
c^2}\frac{\lambda_i+\lambda_e}{\lambda_i}\nabla\psi \nonumber\\
& &\\
& &{\bf{\omega}}_i^\dagger={\bf{\omega}}_i-\frac{2\pi n q^2}{m_i
c^2}\frac{\lambda_i+\lambda_e}{\lambda_i}\nabla\chi \nonumber
\end{eqnarray}

\noindent In terms of these newly defined variables, the EL
equations can be written as,

\begin{eqnarray}
&
&(\nabla\times)^2{\bf{B}}^\dagger+\left[\frac{(\lambda_i+\lambda_e)q}{c}+\frac{2\pi
n q}{m_i\lambda_i c}-\frac{\lambda_i}{4 m_i L_i}\left( \frac{q
L_i}{c}+m_i\right)^2\right](\nabla\times){\bf{B}}^\dagger \nonumber\\
& &+\frac{2\pi n q^2}{m_i
c^2}\frac{\lambda_i+\lambda_e}{\lambda_i}{\bf{B}}^\dagger=0
\end{eqnarray}

\begin{eqnarray}
&
&(\nabla\times)^2{\bf{\omega_i}}^\dagger+\left[\frac{(\lambda_i+\lambda_e)q}{c}+\frac{2\pi
n q}{m_i\lambda_i c}-\frac{\lambda_i}{4 m_i L_i}\left( \frac{q
L_i}{c}+m_i\right)^2\right](\nabla\times){\bf{\omega_i}}^\dagger \nonumber\\
& &+\frac{2\pi n q^2}{m_i
c^2}\frac{\lambda_i+\lambda_e}{\lambda_i}{\bf{\omega_i}}^\dagger=0
\end{eqnarray}

\noindent The above equations suggest that the daggered fields can
be expressed as a superposition of two Chandrasekhar-Kendall (CK)
eigenfunctions, i.e.,

\begin{eqnarray}
& &{\bf{B}}^\dagger={\bf{Y}}_1+\alpha{\bf{Y}}_2 \nonumber \\
& &\\
& &{\bf{\omega}}_i^\dagger={\bf{Y}}_1+\beta{\bf{Y}}_2 \nonumber
\end{eqnarray}

\noindent where ${\bf{Y}}_k$'s satisfy the relation,

\begin{equation}
\nabla\times{\bf{Y}}_j=\lambda_j{\bf{Y}}_j
\end{equation}

\noindent The $\alpha$ and $\beta$ are constants and quantifies
the non force-free part in ${\bf{B}}^\dagger$ and
${\bf{\omega}}_i^\dagger$ respectively. We assume that the $x-y$
plane represents the photospheric surface and the arcades are
along the $z$-direction. In cartesian coordinates a convenient
representation for ${\bf{Y}}$ with $y$ symmetry is

\begin{equation}
{\bf{Y}}=\nabla\times(\phi\hat{e}_y)+\frac{1}{\lambda}
\nabla\times\nabla\times(\phi\hat{e}_y)
\end{equation}

\noindent For (26) to be true, $\phi$ should satisfy the Helmholtz
equation

\begin{equation}
\nabla^2\phi+\lambda^2\phi=0
\end{equation}

\noindent With $y$-symmetry $\phi$ is given by,

\begin{equation}
\phi=\cos(x\sqrt{k^2+\lambda^2})\exp[-kz]
\end{equation}

\noindent Since the arcade solutions are bounded only at $z=0$ and
dies out as $z\rightarrow\infty$ \cite{kn:bellanbook,
kn:bellanpaper}, we have assumed an exponentially decaying
solution along the z-direction. Also, we have written only the
$\cos$ part so that the different field components agree with that
given in the works of Browning {\emph{et. al. }}
\cite{kn:browning}. From equations (27) and (29) different
component of the daggered fields may be written as,

\begin{eqnarray}
& &B_x^\dagger=\left[\sqrt{\kappa_1^2-\lambda_1^2}\cos(\kappa_1
x)e^{-z\sqrt{\kappa_1^2-\lambda_1^2}}+\alpha
\sqrt{\kappa_2^2-\lambda_2^2}\cos(\kappa_2
x)e^{-z\sqrt{\kappa_2^2-\lambda_2^2}}\right] \nonumber\\
& &B_y^\dagger=\left[\lambda_1\cos(\kappa_1
x)e^{-z\sqrt{\kappa_1^2-\lambda_1^2}}+\alpha
\lambda_2\cos(\kappa_2
x)e^{-z\sqrt{\kappa_2^2-\lambda_2^2}}\right] \\
& &B_z^\dagger=-\left[\kappa_1\sin(\kappa_1
x)e^{-z\sqrt{\kappa_1^2-\lambda_1^2}}+\alpha \kappa_2\sin(\kappa_2
x)e^{-z\sqrt{\kappa_2^2-\lambda_2^2}}\right] \nonumber
\end{eqnarray}

\noindent and

\begin{eqnarray}
&
&\omega_x^\dagger=\left[\sqrt{\kappa_1^2-\lambda_1^2}\cos(\kappa_1
x)e^{-z\sqrt{\kappa_1^2-\lambda_1^2}}+\beta
\sqrt{\kappa_2^2-\lambda_2^2}\cos(\kappa_2
x)e^{-z\sqrt{\kappa_2^2-\lambda_2^2}}\right] \nonumber\\
& &\omega_y^\dagger=\left[\lambda_1\cos(\kappa_1
x)e^{-z\sqrt{\kappa_1^2-\lambda_1^2}}+\beta \lambda_2\cos(\kappa_2
x)e^{-z\sqrt{\kappa_2^2-\lambda_2^2}}\right] \\
& &\omega_z^\dagger=-\left[\kappa_1\sin(\kappa_1
x)e^{-z\sqrt{\kappa_1^2-\lambda_1^2}}+\beta \kappa_2\sin(\kappa_2
x)e^{-z\sqrt{\kappa_2^2-\lambda_2^2}}\right] \nonumber
\end{eqnarray}

\noindent where

\begin{equation}
\kappa_j^2=\lambda_j^2+k_j^2~~~~~~~~~~~~~~~~~~~~~j=1,~2
\end{equation}

\noindent The gradient part in equation (22) can be calculated by
solving (21) in the region described by $z>0$.

\begin{equation}
\psi(\chi)=E(F)\sin(lx)e^{-lz}
\end{equation}

\noindent Note that we have once gain assumed an exponentially
decaying solution along the z-direction to be in conformity with
the arcade structures. From equation (22), different field
components are obtained as,

\begin{eqnarray}
& &B_x=\left[\sqrt{\kappa_1^2-\lambda_1^2}\cos(\kappa_1
x)e^{-z\sqrt{\kappa_1^2-\lambda_1^2}}+\alpha
\sqrt{\kappa_2^2-\lambda_2^2}\cos(\kappa_2
x)e^{-z\sqrt{\kappa_2^2-\lambda_2^2}}+E\cos(lx)e^{-lz}\right] \nonumber\\
& &B_y=\left[\lambda_1\cos(\kappa_1
x)e^{-z\sqrt{\kappa_1^2-\lambda_1^2}}+\alpha
\lambda_2\cos(\kappa_2
x)e^{-z\sqrt{\kappa_2^2-\lambda_2^2}}\right] \\
& &B_z=-\left[\kappa_1\sin(\kappa_1
x)e^{-z\sqrt{\kappa_1^2-\lambda_1^2}}+\alpha \kappa_2\sin(\kappa_2
x)e^{-z\sqrt{\kappa_2^2-\lambda_2^2}}+E\sin(lx)e^{-lz}\right]
\nonumber
\end{eqnarray}

\noindent and

\begin{eqnarray}
& &\omega_x=\left[\sqrt{\kappa_1^2-\lambda_1^2}\cos(\kappa_1
x)e^{-z\sqrt{\kappa_1^2-\lambda_1^2}}+\beta
\sqrt{\kappa_2^2-\lambda_2^2}\cos(\kappa_2
x)e^{-z\sqrt{\kappa_2^2-\lambda_2^2}}+F\cos(lx)e^{-lz}\right] \nonumber\\
& &\omega_y=\left[\lambda_1\cos(\kappa_1
x)e^{-z\sqrt{\kappa_1^2-\lambda_1^2}}+\beta \lambda_2\cos(\kappa_2
x)e^{-z\sqrt{\kappa_2^2-\lambda_2^2}}\right] \\
& &\omega_z=-\left[\kappa_1\sin(\kappa_1
x)e^{-z\sqrt{\kappa_1^2-\lambda_1^2}}+\beta \kappa_2\sin(\kappa_2
x)e^{-z\sqrt{\kappa_2^2-\lambda_2^2}}+F\sin(lx)e^{-lz}\right]
\nonumber
\end{eqnarray}

\noindent where parameters $\kappa_j$, $\lambda_j$ (for $j=1,~2$),
$\alpha$ and $\beta$ are to be determined from the boundary
conditions. In the following, we attempt a simplified solution
without any crossings and knots. The condition for that is
\cite{kn:bellanbook, kn:bellanpaper}

\begin{equation}
k_1=k_2=k=l
\end{equation}

\noindent or equivalently

\begin{equation}
\kappa_1^2-\lambda_1^2=\kappa_2^2-\lambda_2^2=k^2
\end{equation}

\noindent The field components for this case then reduce to

\begin{eqnarray}
& &B_x=\sqrt{\kappa_1^2-\lambda_1^2}\left[\cos(\kappa_1
x)+\alpha\cos(\kappa_2
x)+\frac{E}{\sqrt{\kappa_1^2-\lambda_1^2}}\cos(x\sqrt{\kappa_1^2-\lambda_1^2}
\right]e^{-kz} \nonumber\\
& &B_y=\left[\lambda_1\cos(\kappa_1
x)+\alpha\lambda_2\cos(\kappa_2 x)\right]e^{-kz}
\\
& &B_z=-\left[\kappa_1\sin(\kappa_1 x)+\alpha
\kappa_2\sin(\kappa_2
x)+E\sin(x\sqrt{\kappa_1^2-\lambda_1^2})\right]e^{-kz}\nonumber
\end{eqnarray}

\noindent and the vorticity components are obtained as,

\begin{eqnarray}
& &\omega_x=\sqrt{\kappa_1^2-\lambda_1^2}\left[\cos(\kappa_1
x)+\beta\cos(\kappa_2
x)+\frac{F}{\sqrt{\kappa_1^2-\lambda_1^2}}\cos(x\sqrt{\kappa_1^2-\lambda_1^2}
\right]e^{-kz} \nonumber\\
& &\omega_y=\left[\lambda_1\cos(\kappa_1
x)+\beta\lambda_2\cos(\kappa_2 x)\right]e^{-kz}
\\
& &\omega_z=-\left[\kappa_1\sin(\kappa_1 x)+\beta
\kappa_2\sin(\kappa_2
x)+F\sin(x\sqrt{\kappa_1^2-\lambda_1^2})\right]e^{-kz}\nonumber
\end{eqnarray}

\noindent A point to be mentioned is, as the boundary conditions
(6) or (7) has to be obeyed at all the points on the $x-y$ plane,
we have no other option but to assume that this condition is
satisfied automatically. In other words we are assuming the
photospheric fields are to be of the same form as the above set of
equations with $z=0$. In the above set for ${\bf{B}}$ if we put
$\alpha=E=0$, the corresponding magnetic field represents the
photospheric boundary conditions used by Finn \emph{et. al.}
\cite{kn:guzdar}. Here, we are assuming a more general but
structurally similar photospheric boundary condition. Equations
(6) or (7) then do not yield any condition through which different
parameters can be determined. To establish the possibility that
the MDR relaxed state does support meaningful flow-containing non
force-free arcade structures, in the following we consider the
simplest two-dimensional non force-free state described by the
condition,

\begin{equation}
\lambda_1=-\lambda_2=\lambda
\end{equation}

\noindent For such a choice, different field components can be
written as,

\begin{eqnarray}
& &B_x=\sqrt{\gamma^2-1}\left[(1+\alpha)\cos(\gamma
x)+\frac{E}{\sqrt{\gamma^2-1}}\cos(x\sqrt{\gamma^2-1})\right]e^{-z\sqrt{\gamma^2-1}}
\nonumber\\
& &B_y=\left(1-\alpha\right)\cos(\gamma x)e^{-z\sqrt{\gamma^2-1}}
\\
& &B_z=-\left[(1+\alpha)\gamma\sin(\gamma
x)+E\sin(x\sqrt{\gamma^2-1})\right]e^{-z\sqrt{\gamma^2-1}}
\nonumber
\end{eqnarray}

\noindent where

\begin{equation}
\gamma=\frac{\kappa}{\lambda}
\end{equation}

\noindent and $E$, $\kappa$ and $\lambda$ are redefined as

\begin{eqnarray}
E=\frac{E}{\lambda} \nonumber\\
\kappa=\kappa L \\
\lambda=\lambda L \nonumber
\end{eqnarray}

\noindent $L$ being the characteristic length scale representing
the width of the arcade in the $x-y$ plane and $x$ and $z$ are
normalized w.r.t it. The coefficient $\alpha$ is obtained from the
requirement that at $x=1$ the magnetic field lines are open. For
that we need

\begin{equation}
B_x=B_y=0
\end{equation}

\noindent which gives

\begin{eqnarray}
& &\alpha=1\nonumber\\
& &\\
& &E=-\frac{2\sqrt{\gamma^2-1}\cos\gamma}{\cos\sqrt{\gamma^2-1}}
\nonumber
\end{eqnarray}

\noindent With these values for $\alpha$ and $E$, different field
components are obtained as,

\begin{eqnarray}
& &B_x=\sqrt{\gamma^2-1}\left[2\cos\gamma
x-\frac{2\cos\gamma}{\cos\sqrt{\gamma^2-1}}\cos(x\sqrt{\gamma^2-1})\right]
e^{-z\sqrt{\gamma^2-1}}\nonumber\\
& &B_y=0\\
& &B_z=-\left[2\gamma\sin\gamma
x-\frac{2\sqrt{\gamma^2-1}\cos\gamma}{\sin\sqrt{\gamma^2-1}}\sin(x\sqrt{\gamma^2-1})\right]
e^{-z\sqrt{\gamma^2-1}}\nonumber
\end{eqnarray}

\noindent As we have pointed out earlier, the resulting magnetic
field is two-dimensional with $B_y=0$. This is just a simplified
case and more elaborate three-dimensional magnetic fields can be
obtained for other choices of the eigenvalues. But, here our
objective is rather modest; to model and look for arcade-type
magnetic fields as possible MDR states. To do that a
two-dimensional magnetic field would suffice. The third dimension
essentially adds twist to the magnetic field lines and is believed
to be responsible for $S$-shaped solar prominences
\cite{kn:bellanpaper, kn:bellanbook}. We leave this as our future
work. The Arcade structures are obtained by solving the magnetic
field line equation,

\begin{equation}
\frac{dz}{dx}=\frac{B_z}{B_x}
\end{equation}

\noindent and are depicted in figures (1)-(5) for different
$\gamma$ values. It is interesting to observe that along with the
single-arcade structure, the MDR relaxed state also supports
multiple-arcade solutions. The triple-arcade solutions are
particularly interesting as recent observations with the LASCO
coronagraph on board the SOHO spacecraft suggest that the helmet
streamers originating from the sun may have an internal structure
similar to that \cite{kn:schwenn, kn:wiegelmann1}. The observed
triple-arcade structures existed for several days and occasionally
went unstable leading to a new and extraordinarily huge kind of
coronal mass ejection (CME). Although a direct method for
predicting CMEs is not included in our model,  but following
Bellan \cite{kn:bellanbook} a rough qualitative scenario for the
same can be visualized. According to Bellan \cite{kn:bellanbook}
the loss of equilibrium occurs when the altitudinal decay factor
$\gamma$ in equations (46) satisfies the condition

\begin{equation}
\gamma\le 1
\end{equation}

\noindent For $\gamma\le 1$ the magnetic fields no longer decay
with increasing height from the solar surface but are oscillatory
in nature. In the footsteps of Bellan, we may also propose the
following. As $\gamma$ decreases the arcades passes from more
stable to unstable states and as the instability threshold is
crossed, loss of equilibrium occurs with erupting prominences. In
other words, this model requires the multiple-arcade structures to
be more stable compared to their single-arcade counterparts and
indeed, in a recent work based on resistive MHD, Wiegelmann
\emph{et al} have shown that triple streamers are usually more
stable than a single streamer \cite{kn:wiegelmann2}. Alongside, we
want to put a word of caution. The argument presented here is only
qualitative and  indicates that the MDR relaxed state has the
potential to incorporate the phenomenon of CME. An actual
stability calculation for the states described by equations (46)
can decide the issue.

Although, equation (20) supports field-aligned flows as a
non-unique solution obtained with a special gauge, in the
following we look for more general solutions. Different flow
components can be calculated by realizing,

\begin{eqnarray}
& &\frac{\partial u_y}{\partial z}=\omega_x \nonumber\\
& &\frac{\partial u_x}{\partial z}-\frac{\partial u_z}{\partial x}=\omega_y \\
& &\frac{\partial u_y}{\partial x}=\omega_z \nonumber
\end{eqnarray}

\noindent where we have dropped the subscript $i$ for convenience.
From the first and the last set of the above equations, $y$
component of the flow-field is obtained as,

\begin{equation}
u_y=\left[(1+\beta)\cos\gamma
x+\frac{F}{\sqrt{\gamma^2-1}}\cos(x\sqrt{\gamma^2-1}\right]e^{-z\sqrt{\gamma^2-1}}
\end{equation}

\noindent Other two flow components are obtained from the second
of the equations (49), which gives

\begin{equation}
\frac{\partial u_x}{\partial z}-\frac{\partial u_z}{\partial x}
=(1-\beta)\cos\gamma x e^{-z\sqrt{\gamma^2-1}}
\end{equation}

\noindent Assuming an exponentially decaying form for the flows,
the flow components in $x$ and $z$ direction can easily be
calculated from the above equation and is obtained as,

\begin{eqnarray}
&
&u_x=\frac{\beta-1}{(\sqrt{\gamma^2-1}-\gamma)\cos\sqrt{\gamma^2-1}}
\left[\cos\sqrt{\gamma^2-1}\cos(\gamma
x)-\cos\gamma\cos(x\sqrt{\gamma^2-1})\right]e^{-z\sqrt{\gamma^2-1}} \nonumber\\
& & \\
 &
&u_z=-\frac{(\beta-1)}{\cos\sqrt{\gamma^2-1}(\sqrt{\gamma^2-1}-\gamma)}
\left[\cos\sqrt{\gamma^2-1}\sin(\gamma x)-\cos\gamma
\sin(x\sqrt{\gamma^2-1})\right]e^{-z\sqrt{\gamma^2-1}}\nonumber
\end{eqnarray}

\noindent To obtain $\beta$ and $F$ we need additional boundary
conditions on flow, which are not available at this time. Actually
by demanding $u_y$ to be zero at the edge, one of the constants
may be calculated but the other remains still unspecified. Also,
we have to look about the matter that in our calculations we have
obtained $B_y=0$ with a non-zero $u_y$. In solar physics it is
hypothesized that due to the shear flow (the y-component) the
footprints of the arcades move on the $x-y$ plane. If so, we must
generate a $3D$ magnetic field with non-zero y-component. Then the
choice $\lambda_1=-\lambda_2$ will not do and we have to look for
more general relations, in other words more general boundary
conditions are required.

Figure (6) shows the $x$-variation of $u_z$ for single-arcade
solution with $\gamma=1.5$. The direction flips sign at the summit
of the arcade and hence resembles an evershed flow. The
corresponding flow lines are depicted in figure (7).

\section{Conclusions}
In summary, in this work we have proposed a formulation suitable
to obtain externally driven relaxed states in two-fluid plasma.
The final relaxed state is obtained by utilizing the principle of
\emph{minimum dissipation rate} (MDR), which is appropriate for
such systems. One novel feature of this formalism is the presence
of non-trivial flow field coupling that is absent in the
corresponding single-fluid MDR states. The Euler-Lagrange
equations obtained are double-curl in nature and represent non
force-free configuration. For a constant gauge this reduces to the
double curl Beltrami equation obtained by Mahajan and Yoshida
\cite{kn:mahajan1, kn:mahajan2}. Since the double-curl Beltrami
equation represents one of the steady states of the ideal plasma,
this highlights the possible role of the non-zero plasma flow in
securing a steady, as well as, MDR relaxed state.

To establish the plausibility of MDR principle in predicting
arcade structures, an attempt has been made to solve the
Euler-Lagrange equations in terms of Chadrasekhar-Kendall
eigenfunctions subject to the arcade geometry in cartesian
coordinates with axisymmetry. A solution is obtained for the
two-dimensional case characterized by $B_y=0$. One interesting
feature supported by this new relaxed state is the prediction of
continuous transition from the single arcades to multiple arcade
type solutions with increase in the eigenvalue $\gamma$. A
particular case of interest is the observation of triple-arcade
structures as the MDR relaxed state. The importance of such
structures can be realized from the recent LASCO observations on
SOHO spacecraft that suggest the helmet streamers originating from
the sun may have an internal triple-arcade structure
\cite{kn:wiegelmann1, kn:schwenn}. Also a hypothetical picture of
CME can be postulated where the arcades transit from multiplets to
singlets before the instability threshold $\gamma=1$ is achieved.
This is consistent with the recent findings that the triple
streamers are more stable than the single ones
\cite{kn:wiegelmann2}. In addition, the relaxed state predicts
self-consistent plasma flow, which for the single-arcade solutions
resembles the evershed type.

The above findings definitely points out that the MDR relaxed
states applied to astrophysical plasmas is a worthy case for
further investigations. Any such investigation should involve all
the three components of the magnetic field. Only then it will be
possible to study the effects of twist on the field lines. In
principle, this should generate knots and further studies can be
made on the predictability of $S$-shaped solar prominences.

\acknowledgements{The authors R. Bhattacharyya and M. S. Janaki
are extremely thankful to Prof. Parvez N. Guzdar for his
encouragements and helpful discussions during the initiation of
this work.}

\newpage
\begin{figure}[h]
\includegraphics{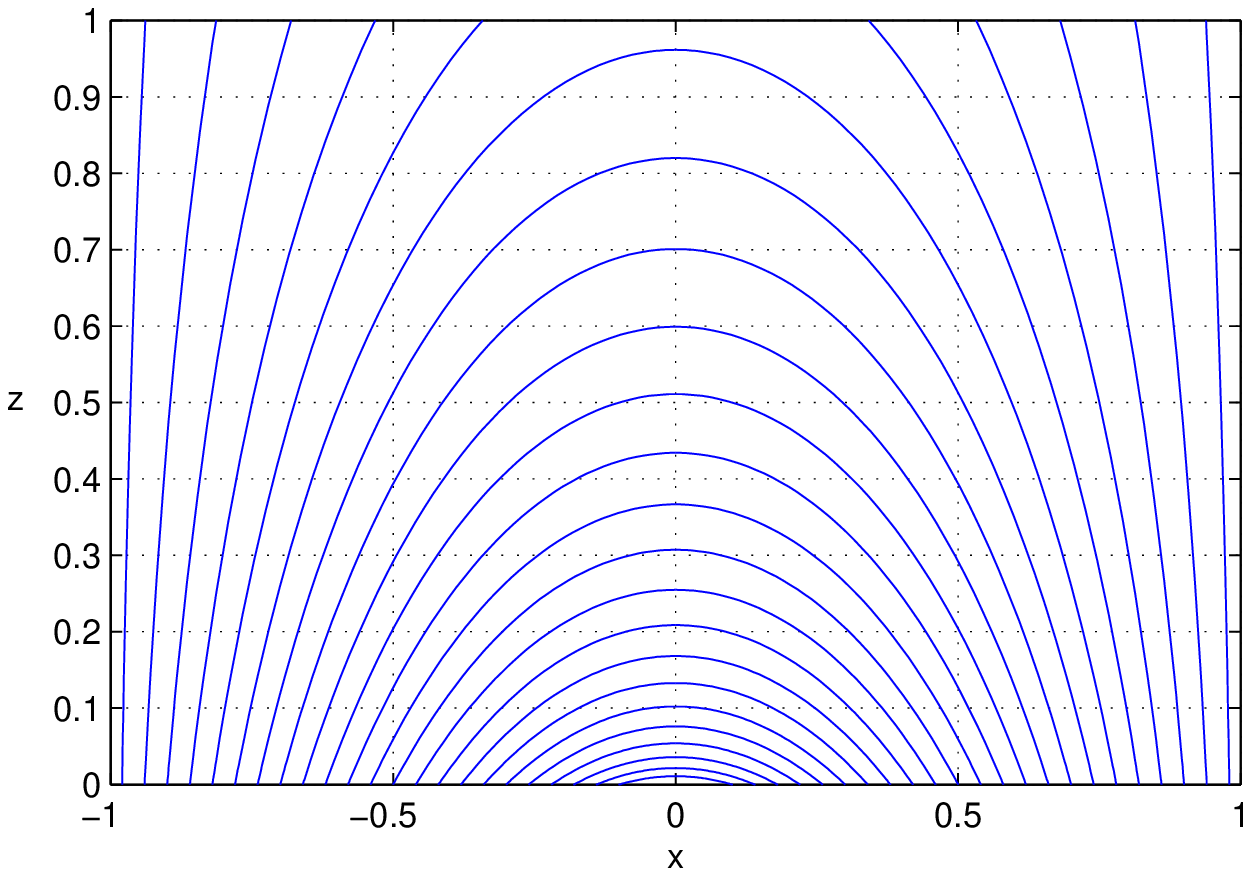}
\caption{\label{fig:1} Magnetic field lines corresponding to
single arcade structures. $\gamma=1.5$}
\end{figure}

\begin{figure}
\includegraphics{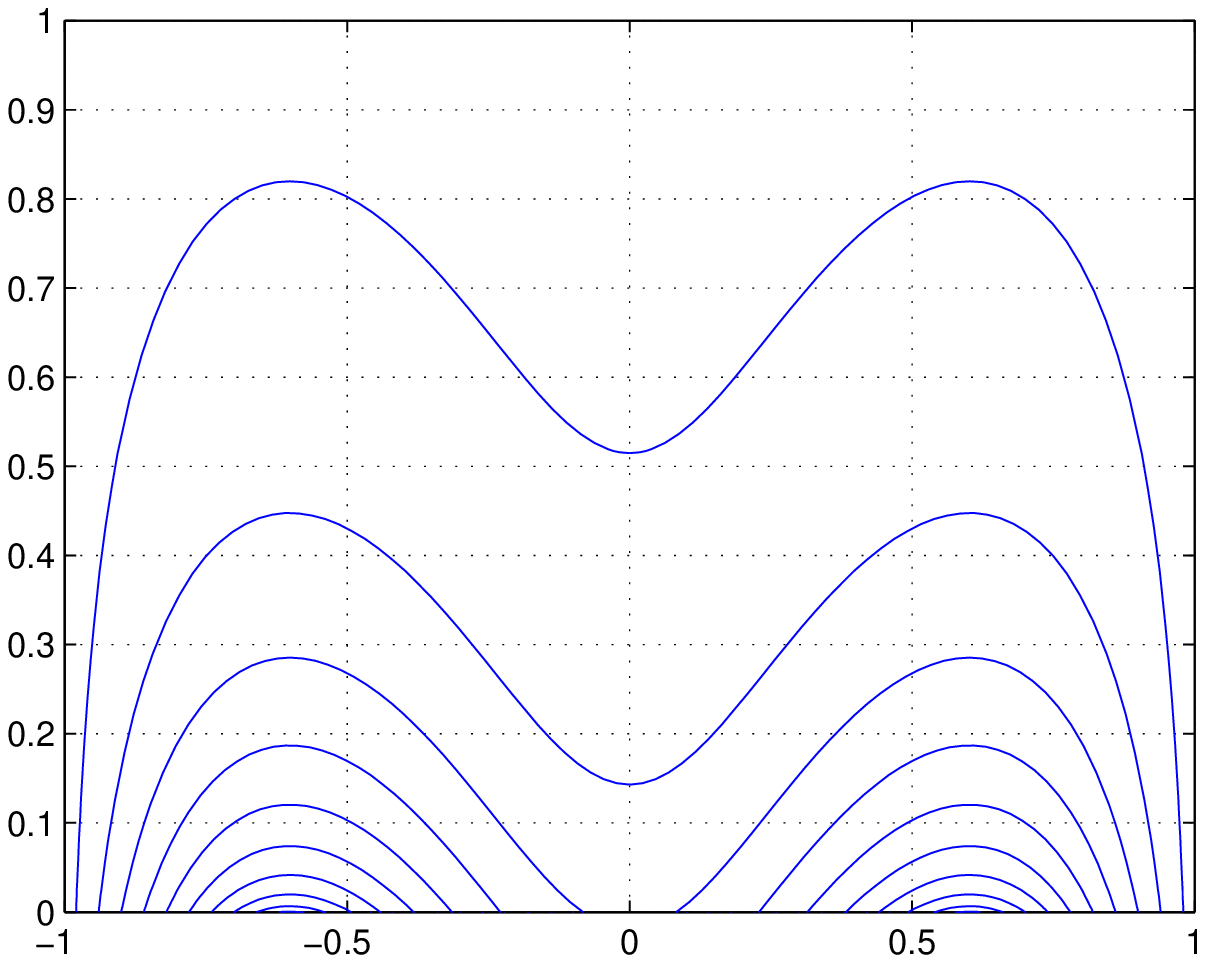}
\caption{\label{fig:2} Magnetic field lines corresponding to
intermediate  structures between single and double arcades.
$\gamma=3.0$}
\end{figure}

\begin{figure}
\includegraphics{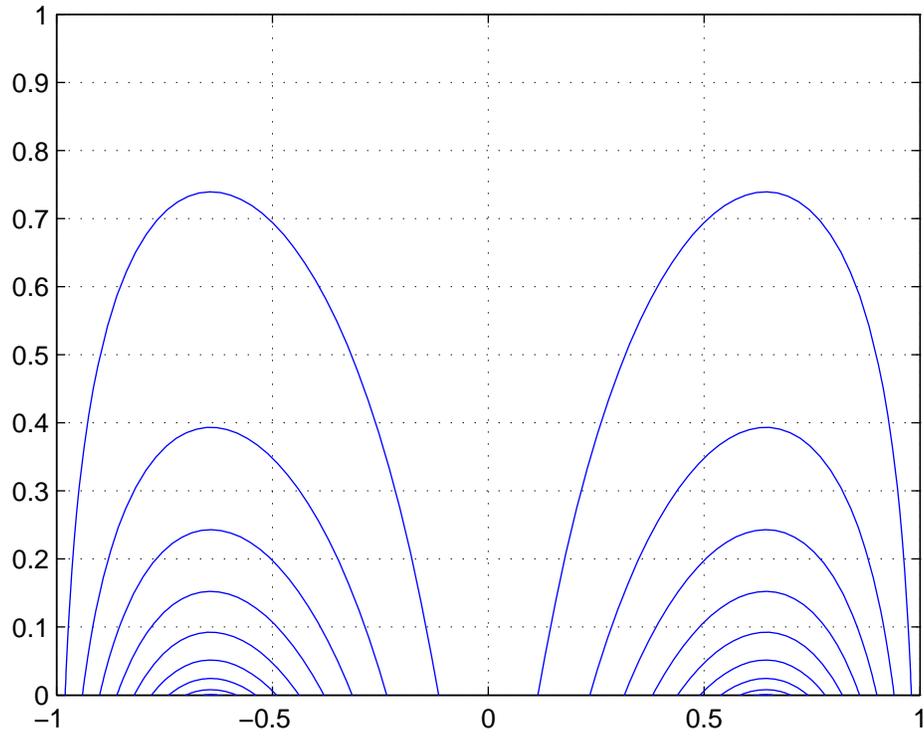}
\caption{\label{fig:3} Magnetic field lines corresponding to
double arcade structures. $\gamma=3.2$}
\end{figure}

\begin{figure}
\includegraphics{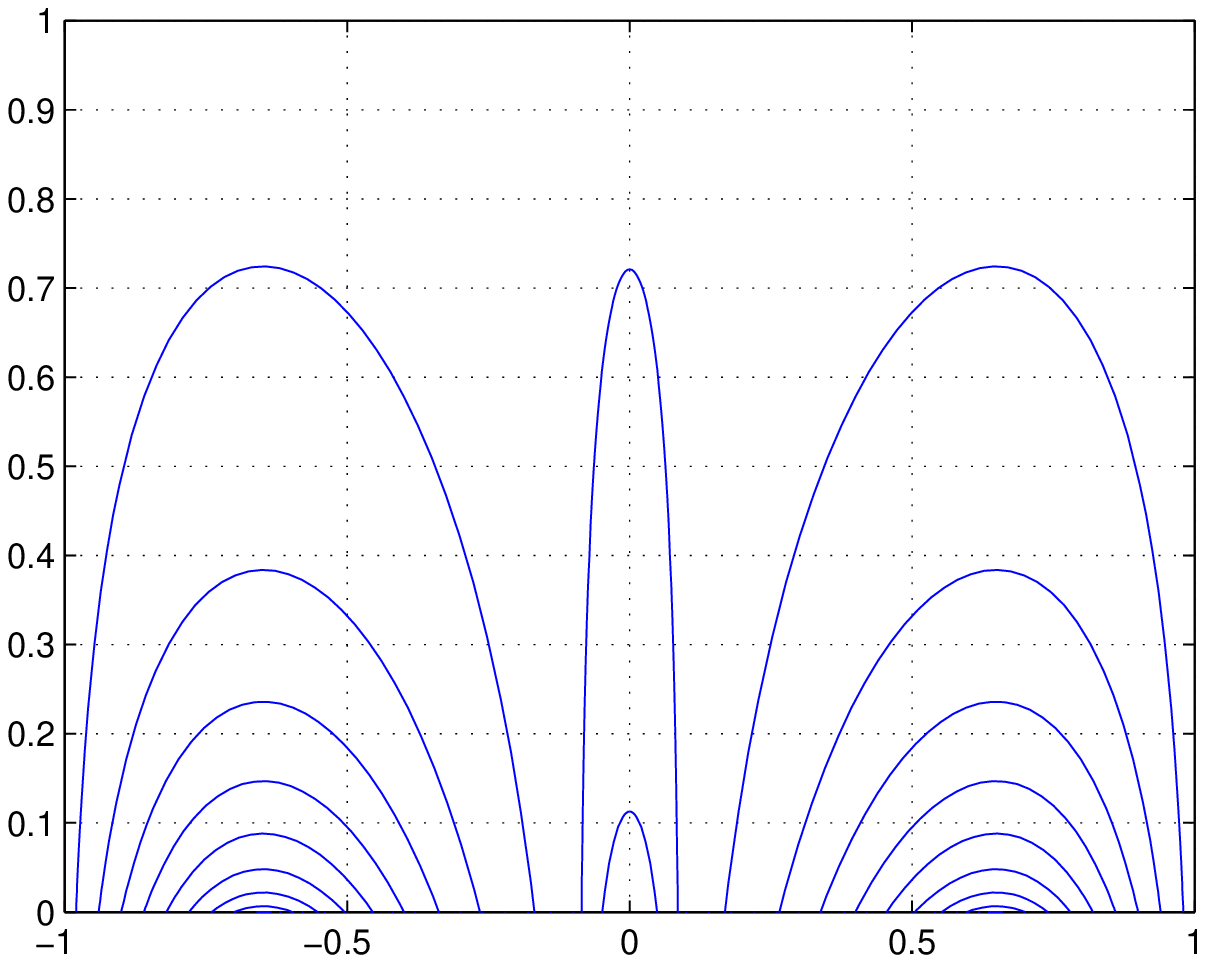}
\caption{\label{fig:4} Magnetic field lines corresponding to
intermediate structures between single and double arcade.
$\gamma=3.247$}
\end{figure}

\begin{figure}
\includegraphics{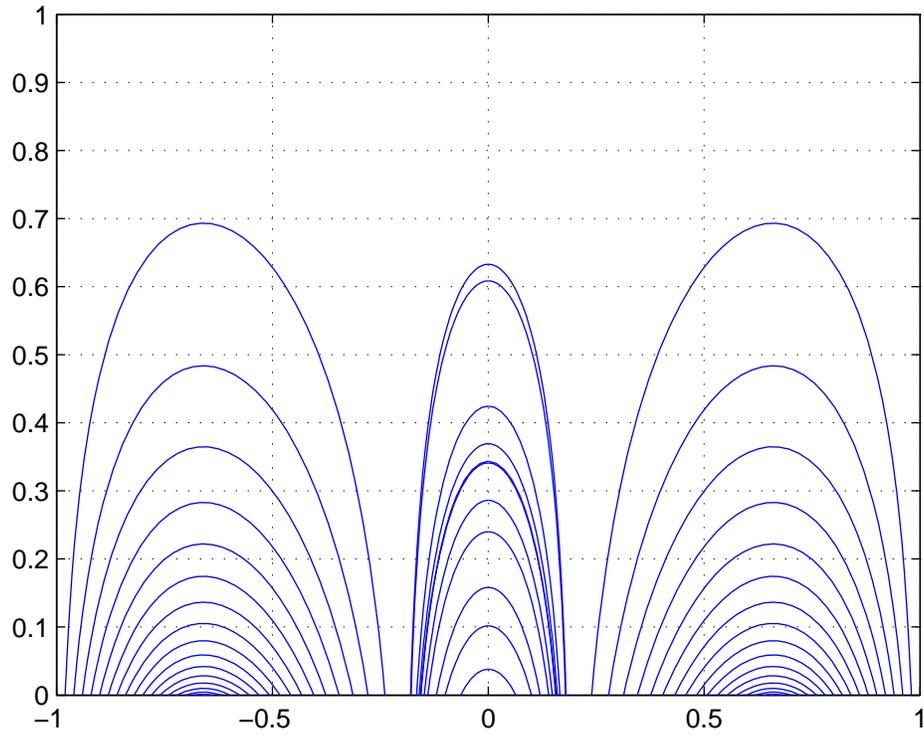}
\caption{\label{fig:5} Magnetic field lines corresponding to
triple arcade structures. $\gamma=3.36$}
\end{figure}

\begin{figure}
\includegraphics{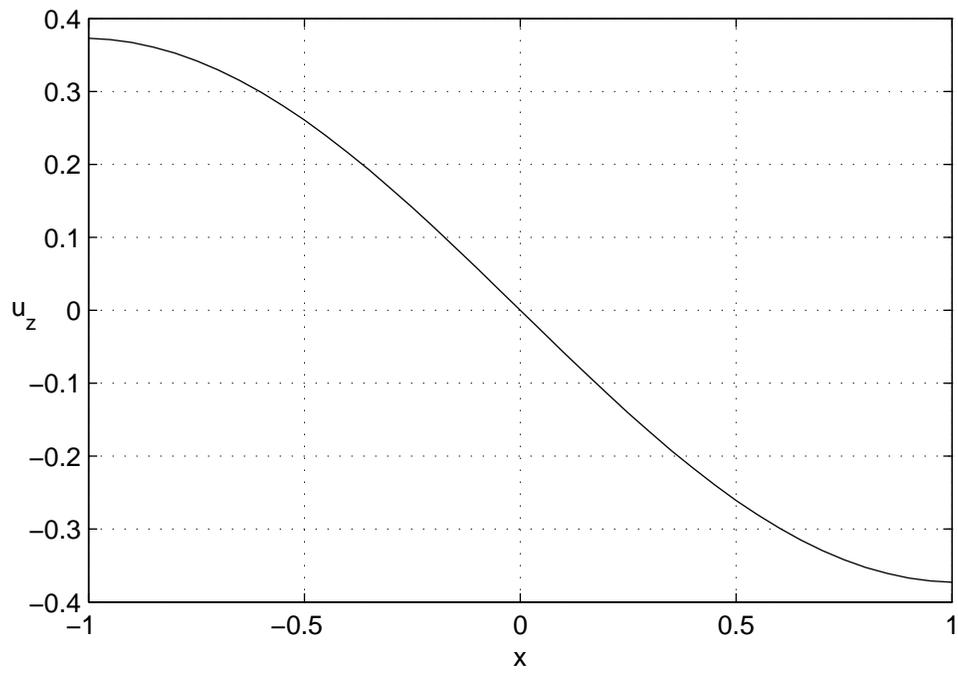}
\caption{\label{fig:6} $u_z$ profile for $\gamma=1.5$}
\end{figure}

\begin{figure}
\includegraphics{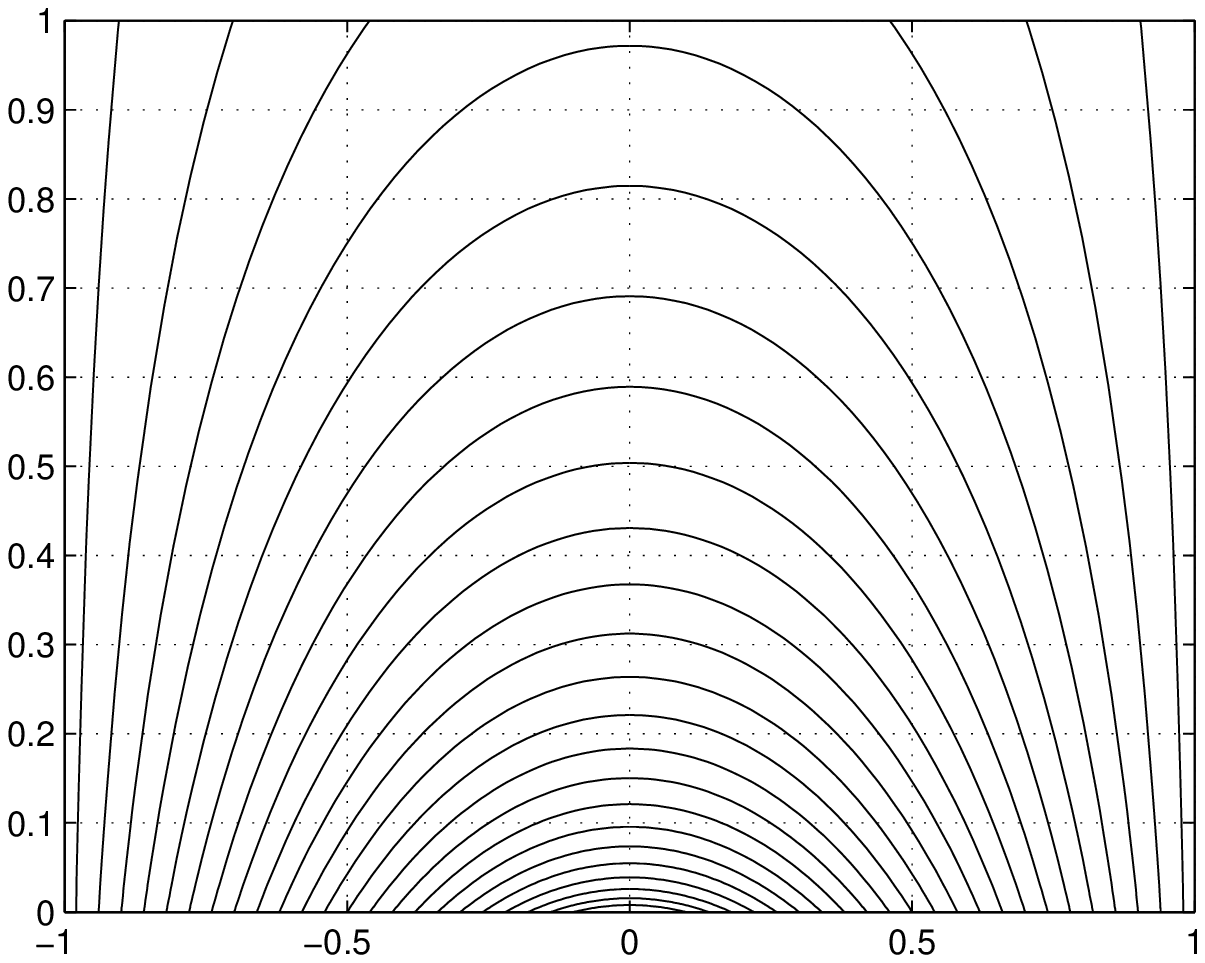}
\caption{\label{fig:7} Flow lines lines corresponding to single
arcade structures. $\gamma=1.5$}
\end{figure}

\end{document}